\documentclass[secnumarabic,nobibnotes,amssymb, aps, showpacs,preprint]{revtex4-1}

\usepackage{graphicx}
\usepackage{amsmath}
\usepackage{amssymb}

\usepackage[]{caption}
\usepackage[caption = false, labelfont= bf]{subfig}

\usepackage{tabularx}
\usepackage{color}
\usepackage{xspace}

\usepackage[squaren,Gray]{SIunits}

\newcommand \Hc{$H_{c2}$\xspace}

\newcommand{\Tsc}{$T_{sc}$\xspace}
\newcommand{\Tc}{$T_{Curie}$\xspace}
\newcommand{\bH}{$\textbf{H}$\xspace}

\newcommand{\xb}{$\textbf{b}$\xspace}
\newcommand{\xa}{$\textbf{a}$\xspace}
\newcommand{\xc}{$\textbf{c}$\xspace}

\begin{document}

\title{Vortex liquid phase in the p-wave ferromagnetic superconductor UCoGe}%

\author{Beilun Wu$^{1,2}$, Dai Aoki$^{1,3}$, Jean-Pascal Brison$^{1}$}%
\email[Corresponding author: ]{jean-pascal.brison@cea.fr}
\affiliation{$^1$University Grenoble Alpes, CEA, INAC-PHELIQS, F-38000 Grenoble, France}
\affiliation{$^2$present address: LBTUAM, Departamento de F\'{i}sica de la Materia Condensada, Instituto Nicol\'{a}s Cabrera and 
IFIMAC 
Universidad Aut\'{o}noma de Madrid, Spain}
\affiliation{$^3$IMR, Tohoko University, Oarai, Ibaraki 311-1313, Japan}

\begin{abstract}
The upper critical field for the field along the b-axis of the orthorhombic ferromagnetic superconductor UCoGe has a particular 
S-shape, akin to the re-entrant superconducting phase of URhGe. In order to explore the evolution of the superconducting phase 
under this transverse magnetic field, we report the thermal conductivity and resistivity measurements, revealing a possible field-induced vortex liquid phase, and supporting a field-induced change of the superconducting order parameter. 
\end{abstract}

\pacs{72.15.Eb, 74.20.Rp, 74.25.fc, 74.70.Tx,  75.50.Cc}

\maketitle
\section{Introduction}
The homogeneous coexistence of ferromagnetism (FM) and superconductivity (SC) in the orthorhombic 
strongly correlated systems UGe$_2$, URhGe, and UCoGe\cite{Saxena2000,Aoki2001,Huy2007,AokiFlouquet2014}
has been well established through NMR and muon spectroscopy \cite{Ohta2010,deVisser2009}. 
Both orders emerge from the uranium 5$f$ electrons, and the mere existence of a superconducting phase in the presence of the strong exchange field controlling the FM state, as well as the absence of  Pauli depairing on their very large upper critical field \cite{SheikinPRB2001,LevyNaturePhys2007,Aoki2009}, points to $p$-wave spin-triplet SC, with an 
"equal-spin-pairing" (ESP) state. 
URhGe and UCoGe share the same $Pnma$ crystal structure, and both show superconductivity at ambient pressure. 
They are weak ferromagnets with the spontaneous magnetization along the \xc-axis. 

A remarkable property of these ferromagnetic superconductors is the reentrant superconducting 
phase (RSC) observed in URhGe\cite{Levy2005}. When the magnetic field is applied along (\bH//\xb), transverse to the easy magnetization axis,
two SC phases are revealed: SC is first suppressed at around \unit{3}{T}, and reappears again at fields around 
\unit{12}{T}, with an even higher transition temperature. The RSC in URhGe has a direct interplay
with a field-induced FM instability at $H_R=$\unit{12}{T}, for which the Curie temperature (\Tc) 
decreases to zero (first order transition) and the magnetic moments reorient completely 
along the applied field direction. Despite the intense experimental and theoretical 
studies\cite{Tokunaga2015,Kotegawa2015,Yelland2011,Gourgout2016,Miyake2008,HattoriTheory2013}, 
it is not known whether the superconducting order parameter has the same symmetry in the low field and in the reentrant phase. 
In the sister compound UCoGe, the situation is similar for the same field direction \bH//\xb: SC is reinforced and the upper critical field (\Hc) has an S-shape \cite{Aoki2009}. Recent work \cite{BraithwaitePRL2018} show that in URhGe, under uniaxial stress applied along the \xb-axis larger than 0.2Gpa, the low field and RSC phases merge into a single phase, as in UCoGe.
Despite the remarkable similarity between the two systems, no spin reorientation has been detected in UCoGe, and the mechanism for the S-shape \Hc also remains unknown, as well as the field-induced evolution of the SC order parameter. The question of field-induced phase transitions is recurrent for triplet superconductors: they have been observed in superfluid $^3$He (A1-phase) \cite{LeggettRMP1975} and UPt$_3$ \cite{JoyntRMP2002}, and in the well-known possible chiral $p$-wave superconductor Sr$_2$RuO$_4$ \cite{MackenzieRMP2003}. Here we show that for UCoGe, there is also strong experimental supports for a deep change of the superconducting order parameter under transverse field.

\section{Experimental}
We report on thermal conductivity ($\kappa$) and resistivity ($\rho$)
measurements in UCoGe under transverse magnetic field \bH//\xb,  
obtained on two single crystals 
grown with the Czochralski pulling method in a tetra-arc furnace. 
Thermal conductivity measurements have been performed on sample $\#1$, 
bar-shaped with the heat current flowing along the \xc-axis (the same as used in 
Ref.\cite{Taupin2014}, labeled S$_{16}^c$, and in Ref.\cite{Wu2017}). It has a residual resistance ratio (RRR) of 16,  
The thermal conductivity measurements were performed down to \unit{150}{mK} 
in a dilution refrigerator, and in magnetic fields \bH//\xb up to \unit{15}{T}. We use the usual 
one-heater-two-thermometer method, and \unit{15}{\mu m} diameter
gold wires, spot-welded on the sample, to realize the thermal links. The temperature 
rise was limited to $\sim 1\%$.  Four-wire ac-resistivity measurements were performed at the same time, through the same gold-wires, 
allowing direct comparison of thermal and charge transport. 
A two-axis Attocube piezo alignment system (a goniometer
$\sim\pm$\unit{3}{\degree}, and a rotator $\sim$\unit{360}{\degree}), has been used to orient in situ
the sample \xb-axis along the magnetic field (with a sensitivity better than \unit{0.05}{\degree}), by optimizing the \Tsc of the resistive transition under field.
The resistivity has also been measured
on a second single crystal (sample $\#2$, RRR $35$) for magnetic fields up to \unit{16}{T}//\xb, in the same geometry.

\begin{figure}[!t]
\begin{center}
\includegraphics[width = 0.7\columnwidth]{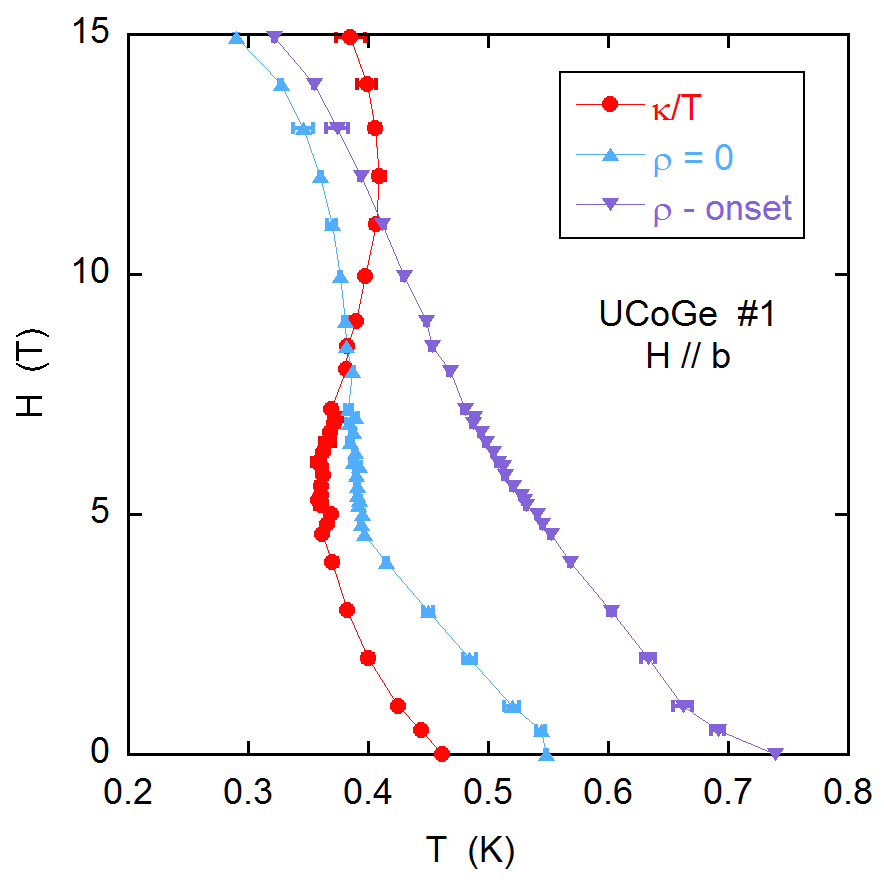}
\caption{
\Hc for \bH//\xb of UCoGe, probed with thermal conductivity (red circles, shown already in Ref.\protect{\cite{Wu2017}}) and resistivity (blue 
triangles for $T_{\rho=0}$ and purple triangles for the onset), measured on sample $\#1$.}
\label{fig-Hc2_Hb}
\end{center}
\end{figure}

\section{Results}
\begin{figure}[!t]
\begin{center}
\subfloat[]{\label{fig-Hb0T}\includegraphics[width = 0.47\columnwidth]{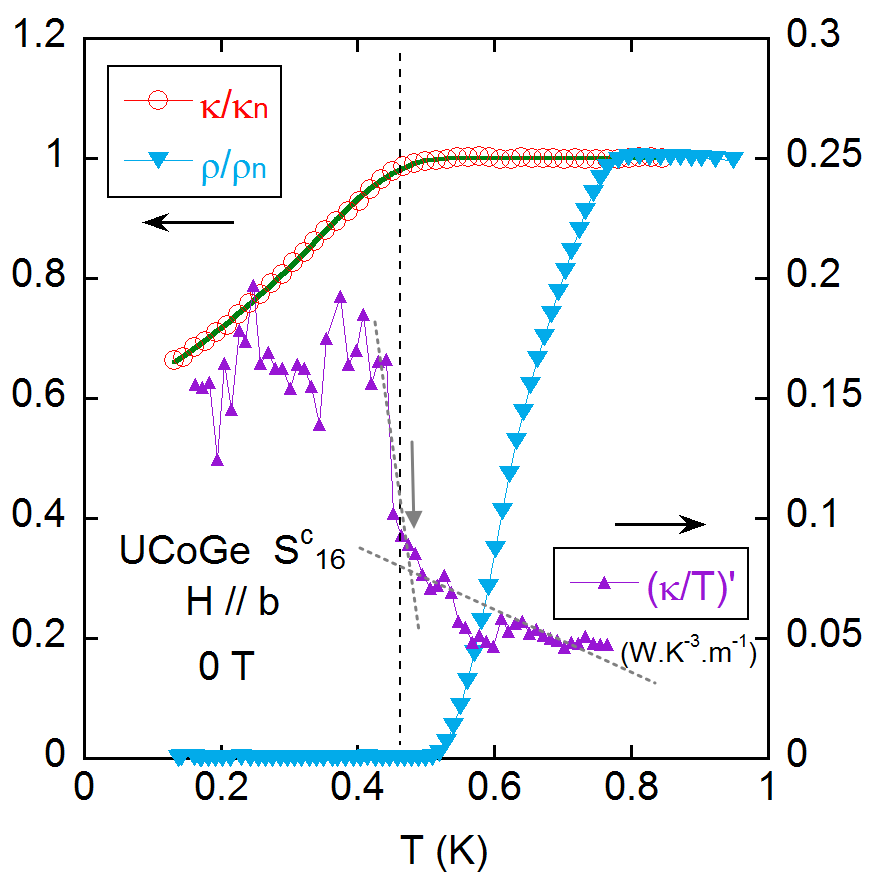}}
\subfloat[]{\label{fig-Hb7T}\includegraphics[width = 0.47\columnwidth]{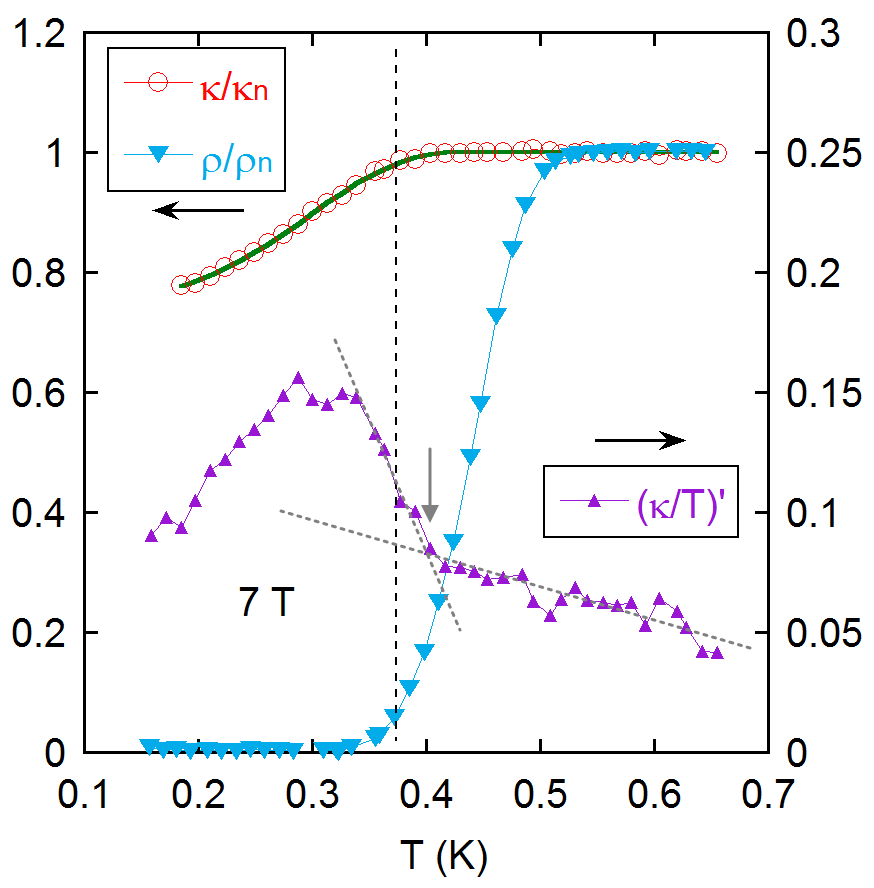}}\\
\subfloat[]{\label{fig-Hb10T}\includegraphics[width = 0.47\columnwidth]{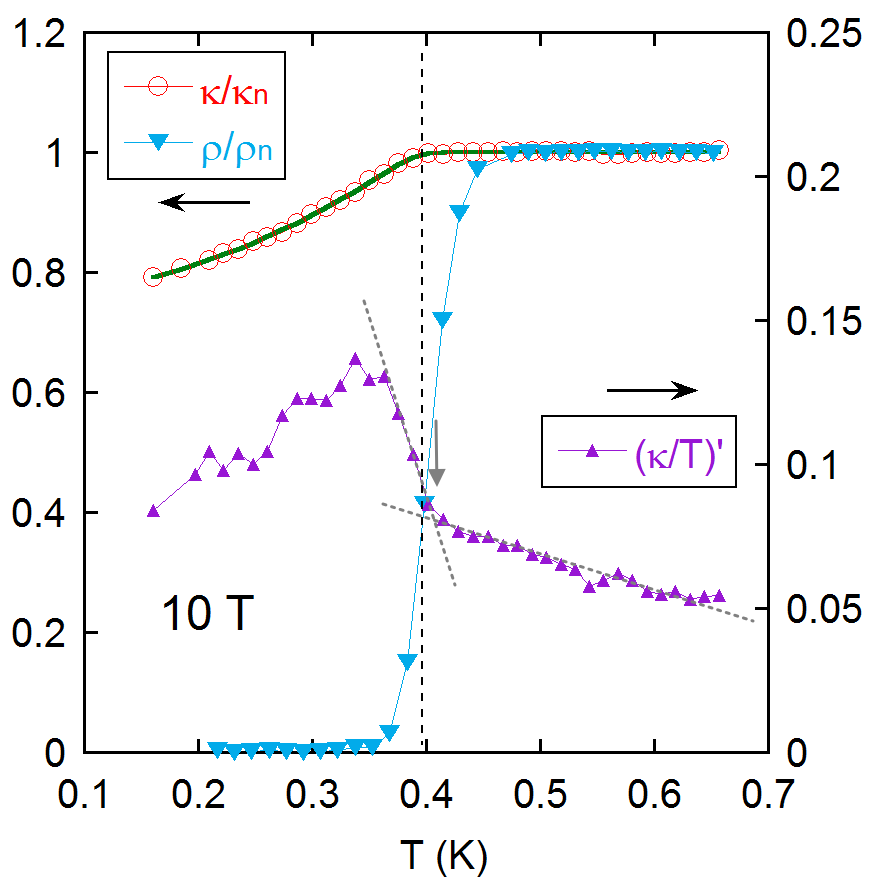}}
\subfloat[]{\label{fig-Hb14T}\includegraphics[width = 0.47\columnwidth]{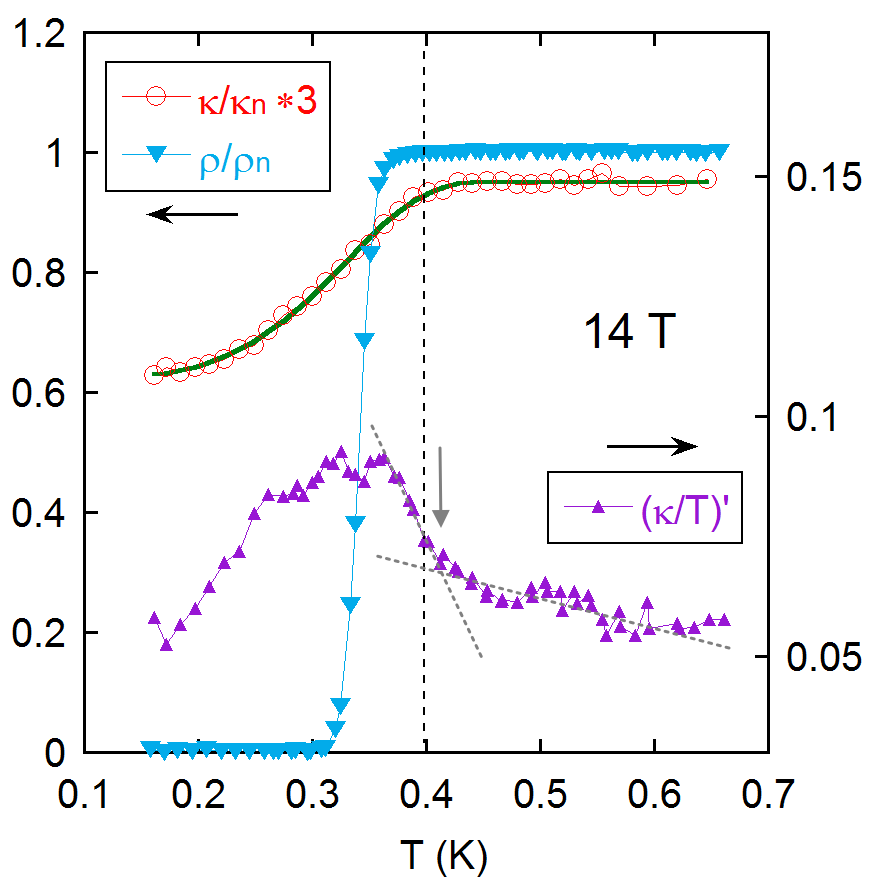}}\\
\caption{\label{fig-deriveHb}
Superconducting transition 
of the sample $\#1$, for different fields \bH//\xb.
(\textbf{(a)}: \unit{0}{T}; \textbf{(b)}: \unit{7}{T}; \textbf{(c)}: \unit{10}{T}; \textbf{(d)}: \unit{14}{T}). 
Red circles: Normalized electronic contribution 
$\kappa/\kappa_n$, with fit (green solid line). 
Blue triangles:  Normalized resistivity ($\rho/\rho_n$).
Purple triangles: Derivative $\partial(\kappa/T)/\partial T$ in \unit{}{J.K^{-3}.m^{-1}} (scale on the right). 
Vertical dashed line: \Tsc obtained from the (green) fit of $\kappa/T$, Gray vertical arrows, \Tsc obtained from $\partial(\kappa/T)/\partial T$. 
}
\end{center}
\end{figure}

Figure \ref{fig-Hc2_Hb} presents  \Hc for \bH//\xc obtained from thermal conductivity 
and resistivity measurements on sample $\#1$. Raw data are presented in the Supplementary Material.  
The rather low sample quality (RRR $= 16$) was chosen on purpose, to get a clear signature of the superconducting 
transition in the thermal conductivity (suppression of the inelastic scattering \cite{Taupin2014}).  
For a precise determination of the transition temperature, $\kappa/T$ data were first analyzed to extract the electronic quasiparticle from the phonon and magnon contributions, and then fitted to extract  \Tsc (same procedure as explained with more details in Ref \cite{Wu2017}). The resistive transitions were also fitted to extract systematically \Tsc from an onset or $\rho=0$ criteria.
Figure \ref{fig-deriveHb} presents the SC transition on the thermal conductivity ($\kappa_{elect}/\kappa_n$), and on the resistivity ($\rho/\rho_n$) for selected magnetic fields. Below $\sim$\unit{8}{T}, the resistivity transition occurs at a temperature higher than the (bulk) superconducting transition on the thermal conductivity, for both the onset and the $\rho=0$ criteria. This is usual, as any
filamentary superconducting path in the sample can induce a resistivity drop, before the bulk superconductivity occurs. What is less usual, but well known for this system, is the fact that the resistive \Tsc can be much larger than the bulk \Tsc, (almost a factor 2 for the onset criterium). A side effect of this very large resistive transition width is that the temperature dependence of \Hc deduced from the resistivity is strongly criterium-dependent, and is also known to be sample dependent (notably the "S-shape"). This was notably a major motivation for a bulk determination of \Hc \cite{Wu2017}.

By contrast, a new phenomenon appears for fields above $\sim$\unit{8}{T}: the resistive transition starts to overlap the bulk transition, and the $\rho=0$ criterium leads a transition temperature below the bulk transition. For fields above $\sim$\unit{12}{T}, even the onset criterium leads to a resistive \Tsc below the bulk determination. This evolution is clearly seen on the selected raw data presented in Fig.\ref{fig-deriveHb}-c-d.
To show the robustness of this result, the derivative with respect to temperature $(\kappa/T)'$ of the raw data (without normalization) is equally presented in Fig.\ref{fig-deriveHb}: $(\kappa/T)'$ is linear in temperature in the normal state in the presented temperature window, so the bulk SC transition would be marked by a change of slope of $(\kappa/T)'$ (grey arrows in figure \ref{fig-Hc2_Hb}). Although more arbitrary (notably for the noise averaging), this determination is in good agreement - yielding even a slightly higher \Tsc - with that obtained
from the fit on the normalized data (vertical dashed line). It can be clearly observed that the resistive transition (and even its onset) shifts to a temperature lower than the bulk SC transition at \unit{10}{T} and \unit{14}{T}. We also checked that the resistive transition does not shift back to larger \Tsc when decreasing the measurement current:  the crossing of the bulk and resistive transitions is not due to Joule heating. 

When a type II superconductor shows an onset of the resistive transition at a temperature lower than the bulk 
\Tsc, the prime suspect is current-driven vortex motion (flux flow or flux creep). 
This is most easily observed in 2D systems, such as organics and high-$T_c$ cuprates, and in some other superconducting systems, where the 
resistive transition follows, instead of the real superconducting transition, a so-called irreversibility line. 
This line corresponds to the freezing of current-induced vortex motion\cite{Palstra1990, Grissonnanche2014, Belin1999}. 
In these systems, the resistive transition is significantly broadened with increasing magnetic field, and, in some 
cases, well below the bulk $T_c$, it shows a sudden drop: this drop would arise from the "freezing transition", from a dissipative vortex-liquid state, 
to a vortex-lattice or glass-like state, where a much stronger pinning efficiency leads to zero resistance
\cite{Safar1992, Kwok1992, Charalamous1992, Welp1996, Okazaki2008, Koshelev1994, Brandt1989}.

\begin{figure}[!t]
\begin{center}
\subfloat[]{\label{fig-width_rho}\includegraphics[height = 0.45\columnwidth]{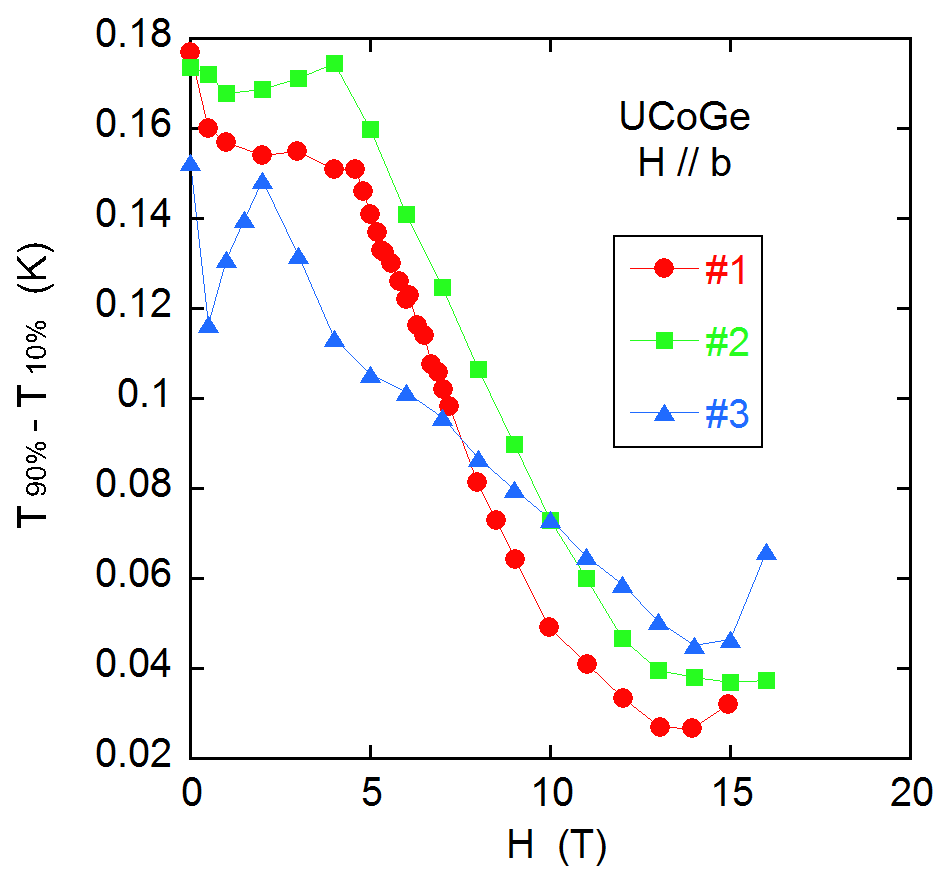}}
\hspace{5pt}
\subfloat[]{\label{fig-Hc2_rho_Hb_Dai}\includegraphics[height = 0.45\columnwidth]{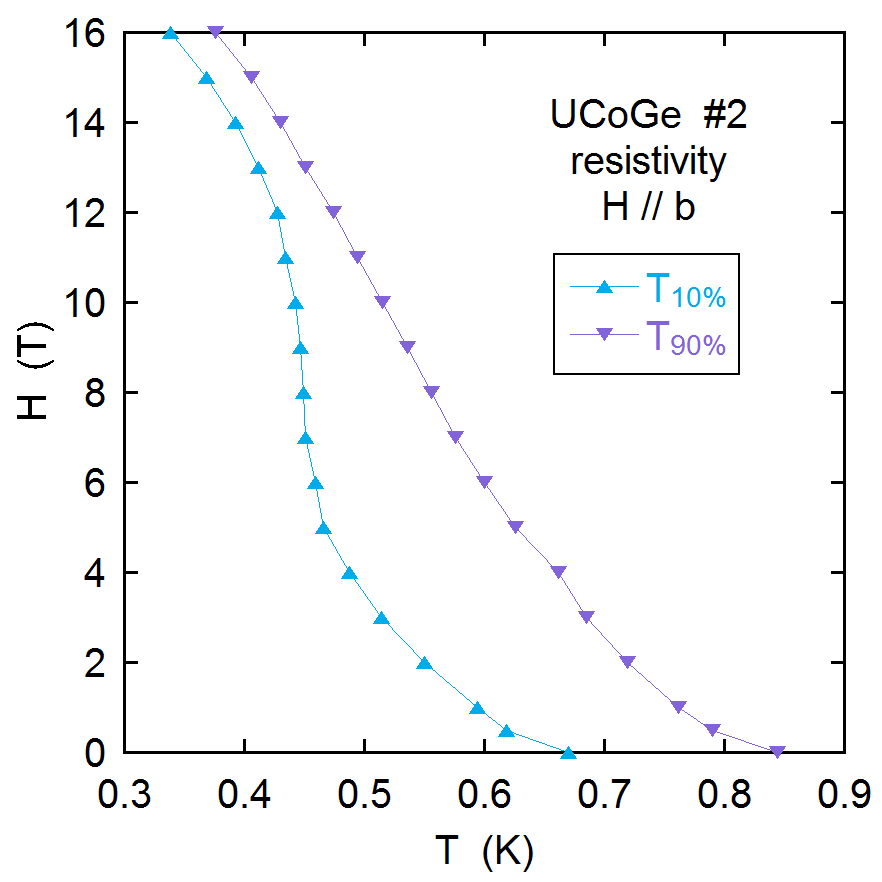}}\\
\caption{\label{fig-UCoGeRhoDai}
\textbf{(a)}: Field dependence of the resistivity transition width, given by $T_{90 \%}-T_{10 \%}$, of UCoGe for \bH//\xb, 
on UCoGe sample $\#1$ and $\#2$, and sample $\#3$ reported in Ref.\cite{Aoki2009}. 
\textbf{(b)}: \Hc of UCoGe for \bH//\xb probed with resistivity on sample $\#2$. 
blue up triangle: $T_{10 \%}$ criterium, purple down triangle: $T_{90 \%}$ criterium. 
}
\end{center}
\end{figure}

In UCoGe, the resistive transitions become much sharper in the high-field region for \bH//\xb. 
This is observed on all UCoGe samples, whatever the exact shape of \Hc for  \bH//\xb (see for example data presented in Ref.\cite{Aoki2009}, referred to as sample $\#3$ in the following, and the raw data shown in the Supplementary Material for both samples $\#1$ and $\#2$ measured in this study).
The same phenomenon is completely absent for the two other field directions \bH//\xa and 
\bH//\xc: in these two cases the resistivity transition enlarges gradually with increasing field, and 
lies always above the bulk \Tsc as determined from thermal conductivity. 
Figure \ref{fig-Hc2_rho_Hb_Dai} shows the onset and the end of the resistive transition for \bH//\xb 
of sample $\#2$: compared to sample $\#1$, \Tsc is larger but the behaviour is the same. Figure \ref{fig-width_rho} shows the field-dependence (\bH//\xb) of the resistive transition 
width of the three samples, defined as $T_{90\%}-T_{10\%}$. 
We can observe  a sudden drop of the transition width, starting at about \unit{5}{T} in samples $\#1$ and $\#2$. 
In sample $\#3$, the evolution of $T_{90 \%}-T_{10 \%}$ is more complicated, 
due to the several steps displayed in its resistive transition, but it is very similar to that of samples $\#1$ and $\#2$ above \unit{5}{T}. 

This evolution of the transition width might have been controlled by the temperature dependence of \Hc itself: for example, the broadening might stop when the slope of \Hc is vertical,
if it originates from a \Tsc distribution leading to an \Hc distribution (slope proportional to \Tsc). But this mechanism cannot lead to a reduction of the broadening. It would, if the broadening corresponded to a distribution of fields inside the sample, but this cannot be the case for fields of order $\sim$\unit{12}{T} in such a strong kappa material. So the most likely origin of the strong reduction of the transition width is a "vortex-freezing transition".

In that sense, UCoGe for \bH//\xb is similar to 
URu$_2$Si$_2$, where the authors of Ref.\cite{Okazaki2008} have observed, with the same measurements, a resistive transition determining a melting transition $T_m$ lower than the bulk \Tsc.
The peculiarities of URu$_2$Si$_2$ (low carrier density system with very heavy effective masses, 1D regime at high fields) put forward in 
\cite{Okazaki2008} as favoring superconducting fluctuations are certainly also present in UCoGe, which has a comparable Sommerfeld coefficient and even larger effective mass (according to the the slope of \Hc at \Tsc for \bH//\xb or \bH//\xa. The most striking difference between the two systems lies in the fact that, while in the tetragonal system URu$_2$Si$_2$ 
the separation of $T_m$ and $T_c$ appears in the whole field range both for the in-plane and perpendicular field direction, 
in UCoGe such a behavior appears only in the high field region for \bH//\xb(above around \unit{8}{T}), as if the pinning mechanism was changing under field. The reinforcement of the pairing mechanism under field \cite{Wu2017}, leading to larger effective masses, larger \Tsc, smaller coherence length, goes in the right direction. But quantitatively, the effect on the critical fluctuation or the melting line as proposed in \cite{Okazaki2008} is certainly not enough to explain the occurence of the vortex liquid phase only at high fields. Another possibility could be a field-induced change in the vortex cores triggered by some change of the $p$-wave order parameter, which could lead to a reduction of the pinning force. Before discussing vortex structures in $p$-wave superconductors, let us first address the question of a field-induced transition in the superconducting state of UCoGe.

\begin{figure}[!t]
\begin{center}
\includegraphics[width = 0.8\columnwidth]{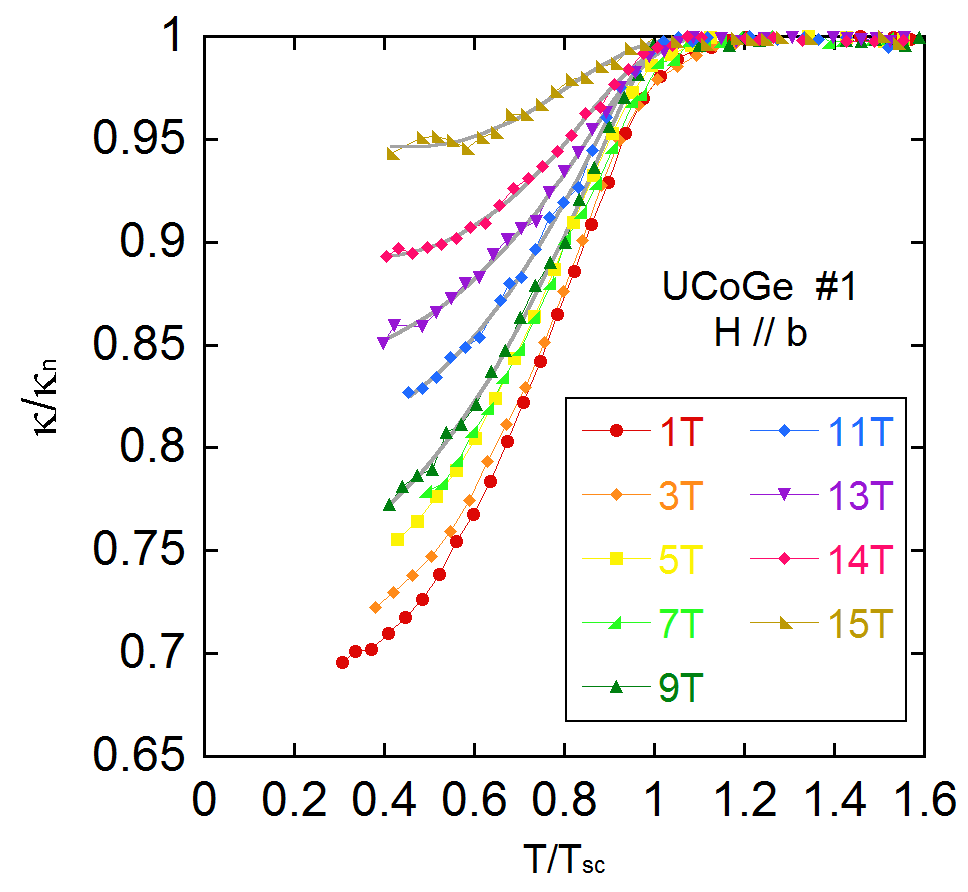}
\caption{\label{fig-KKn_Hb} Electronic contributions to the thermal conductivity normalized to its normal phase 
value $\kappa/\kappa_n$, as a function of temperature normalized to
$T_{sc}(H)$, for fields along the \xb-axis (sample $\# 1$). 
Solid lines are the fits used to determine \Hc (see Ref.\protect{\cite{Wu2017}}).
}
\end{center}
\end{figure}

To study the evolution of the SC under magnetic field, we have further analyzed the $\kappa/T$ data. 
Figure \ref{fig-KKn_Hb} presents the electronic contribution to the thermal conductivity 
normalized to the normal phase values ($\kappa/\kappa_n$), as a function of the normalized temperature ($T/T_{sc}$), at different 
magnetic fields along the \xb-axis. 
For fields between \unit{5}{T} and \unit{9}{T}, $\kappa/\kappa_n$ seems to be almost field independent. 
However, for fields above \unit{9}{T}, $\kappa/\kappa_n$ increases steadily with field and quickly approaches $1$, even though \unit{15}{T} seems still far from $H_{c2}(0)$. This can be quantified by reporting the extrapolated residual thermal conductivity ($\kappa/\kappa_n(T=0)$) versus $H/H_{c2}(0)$. In UCoGe, because the pairing mechanism appears to be field dependent, $H_{c2}(0)$ becomes also field dependent (see \cite{Wu2017}): Figure \ref{fig-KKn_0} displays  $\kappa/\kappa_n$ (with quadratic or linear extrapolation to $T=0$), versus magnetic field (Fig \ref{fig-KKn_0}-a) or $H/H_{c2}(0)$ (Fig \ref{fig-KKn_0}-b) calculated according to the model in Ref. \cite{Wu2017}.

\section{Discussion}
\begin{figure}[!t]
\begin{center}
\subfloat[]{\label{fig-KKn_0_Hb}\includegraphics[width = 0.47\columnwidth]{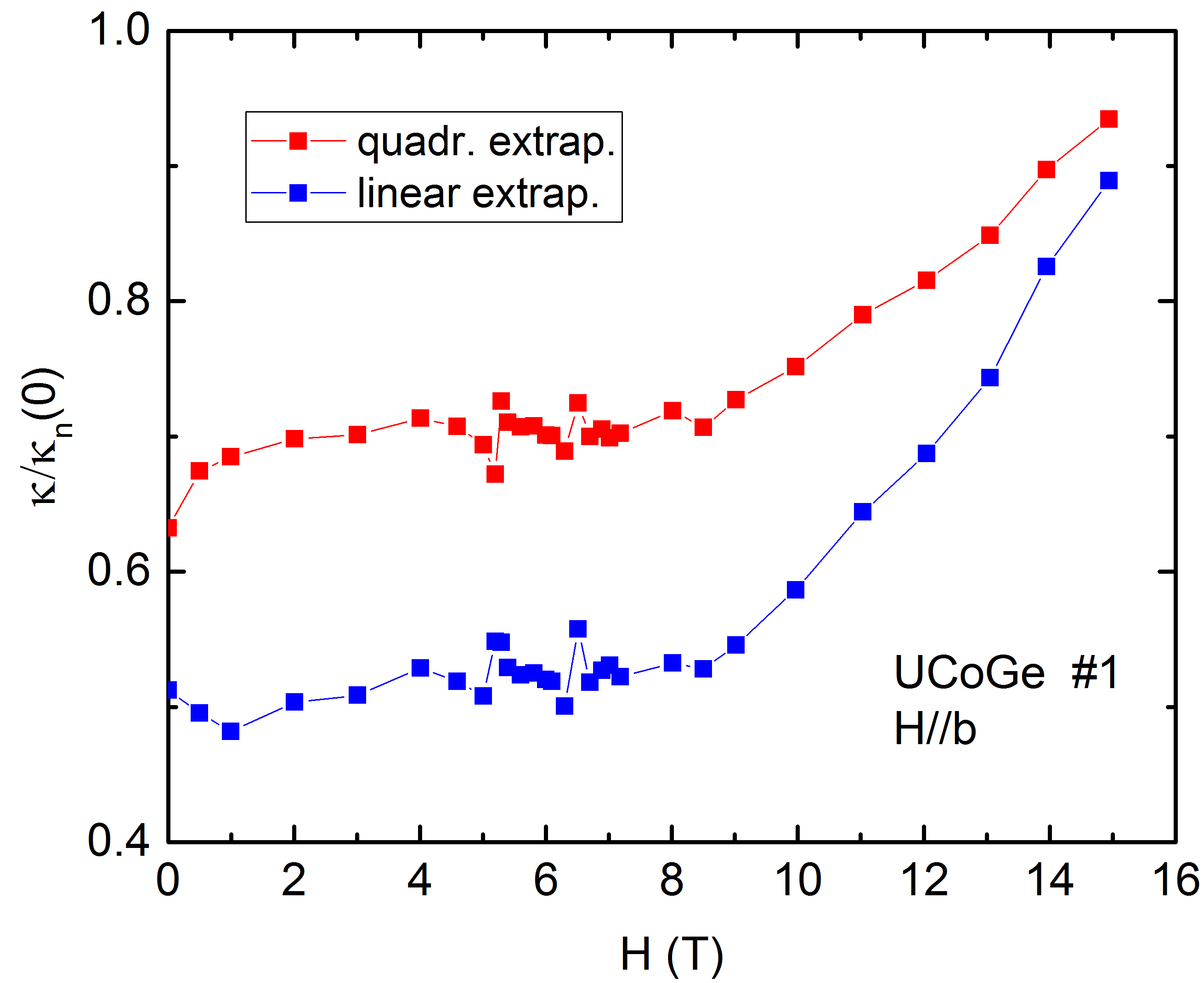}}
\hspace{5pt}
\subfloat[]{\label{fig-KKn_0_Hb_bis}\includegraphics[width = 0.47\columnwidth]{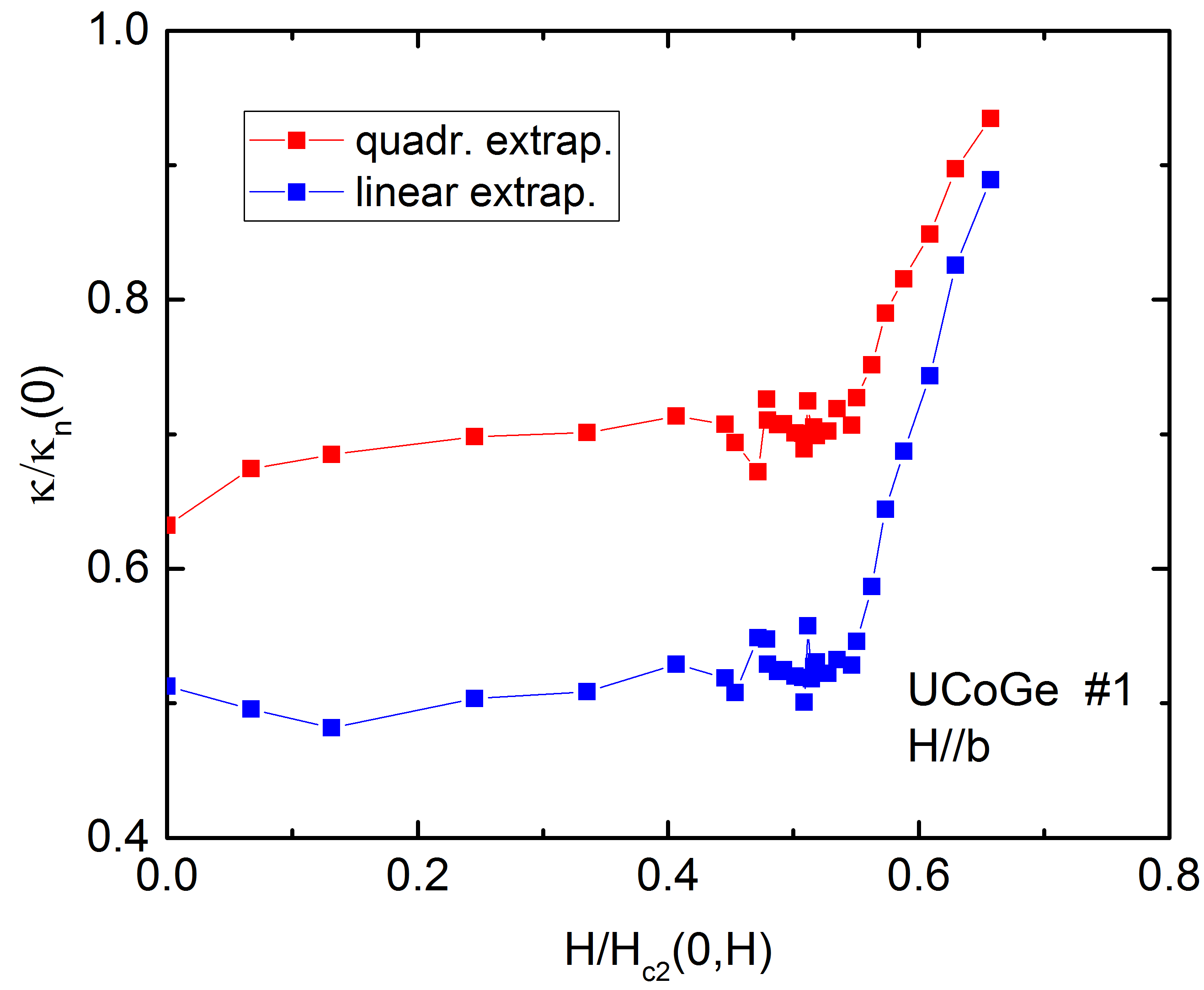}}\\
\caption{\label{fig-KKn_0}
\textbf{(a)}:  $\kappa/\kappa_n(0)$ (sample $\#1$) as a function of field for \bH//\xb in UCoGe. 
\textbf{(b)}:  Same data as a function of $H/H_{c2}(0,H)$, with $H_{c2}(0,H)$ 
obtained from the analyzis in Ref.\protect{\cite{Wu2017}}. Red circles: From quadratic extrapolation to T=0. 
Blue triangles: From linear extrapolation . 
}
\end{center}
\end{figure}

Figure \ref{fig-KKn_0}-(b) shows that, independently from the extrapolation procedure, there is an abrupt increase of the residual thermal conductivity above 9T. Such an increase reflects a change of the electronic quasiparticle excitation spectrum in the superconducting state, leading to a strong enhancement of the residual density of states. This could arise from a field-induced change of the nodal structure of the superconducting order parameter,  for example, due to a change from a line node in the (\xa,\xb) plane of the B-state to a point node along the \xc-axis of the A-state, for an orthorhombic ferromagnetic superconductor \cite{MineevPhysUspek2017}, a situation similar to UPt$_3$ \cite{JoyntRMP2002}; or, alternatively, due to a rotation of the nodal structure with respect to the heat current direction (along the \xc-axis in our case), as a result of a reorientation of the spin quantization axes. Note that a change of the vortex core structure alone is not likely to increase the residual thermal conductivity, owing to the geometry (field perpendicular to the heat current) and to the fact that UCoGe is in the clean limit (mean free path much larger than the coherence length). Careful measurements of \Hc did not reveal (within experimental error) a kink of \Hc, which could have accompanied the transition between two different superconducting states.

There are many factors which could contribute to a field-induced transition inside the superconducting state. First of all, at zero field, UCoGe is generally considered to be in a spin-triplet-ESP state, with a quantization axis along the \xc-axis imposed by the exchange field ($H_{exch}$). $H_{exch}$ is in competition with the external field applied along the \xb axis, and for field values of order 10T, the induced magnetisation along \xb is of the same order than the spontaneous magnetisation in zero field \cite{HuyPRL2008,KnafoPRB2012}. 
Meanwhile, due to the weakening FM order with increasing transverse field along \xb-axis, a reduction of $H_{exch}$ is expected\cite{TadaPRB2016}, resulting in a reduction of the energy gap between the spin-up and spin-down bands. For these reasons, even if no change of the order parameter symmetry occurs, a significant rotation of the d-vector is expected \cite{HattoriTheory2013,TadaPRB2016}, as well as a recovery of a unitary state \cite{TadaPRB2016}. In URhGe, the possibility of a compensation of the exchange field (after the moment rotation along the b-axis) by the external field (Jaccarino-Peter's effect \cite{Jaccarino1962}) has been predicted to lead to non ESP states \cite{HattoriTheory2013}.

Moreover, the nature of the pairing mechanism itself is also influenced by an external field along \xb. At low field, fluctuations are purely longitudinal, along the \xc-axis \cite{HattoriPRL2012}. But in the sister compound URhGe, NMR experiments have shown that ferromagnetic fluctuations also grow along the \xb axis on approaching $H_R$ \cite{Tokunaga2015, Kotegawa2015}. Still in URhGe, uniaxial pressure experiments suggest that fluctuations become more 2D in zero field under stress \cite{BraithwaitePRL2018}, and that a similar effect probably also arises for the RSC phase at $H_R$ . A similar, and expected, field-induced evolution of the pairing mechanism in UCoGe could drive a change of SC order parameter.

These (possible) effects suggest that, when the superconducting state is reinforced under field parallel to \xb, the spin direction of the Cooper pairs might not remain locked to the crystalline axis as it certainly is in low field. This recovered spin degeneracy \cite{HattoriTheory2013} is also favorable to the appearance of non-singular vortices, i.e., vortices where the order parameter does not vanish completely in the core. Such vortices have been predicted to occur for superconductors with multicomponent order parameters (see review \cite{LukyanchukArXiv1995}), either from the orbital degrees of freedom \cite{Tokuyasu1990} or from the spin degrees of freedom in a triplet state  \cite{FujitaJPSJ1994,MineevArXiv2018}, and this field of research on exotic vortices notably in p-wave superconductors \cite{GaraudSciRep2015} or in Bose-Einstein condensate is still very active \cite{KasamatsuPRA2016}. For multicomponent order parameters, it is possible that zeros of the two components of the superconducting order parameter
are located at different points in space, leading to "non-singular" vortex cores (non-zero order parameter in the core). Most of these predictions are valid for isolated vortices (close to $H_{c1}$) \cite{LukyanchukArXiv1995}, but the dissociation of singular vortices in pairs of half quantum vortices \cite{ZhitomirskyJPSJ1995} is also predicted to exist in high fields, in the regime of Abrikosov lattices, for multicomponent (orbital degrees of freedom) order parameters \cite{ZhitomirskyJPSJ1995} or ESP triplet states \cite{ChungNJP2009} (it is even favoured by non-unitary states in this last case). In fact, any ESP $p$-wave state could support half-quantum vortices \cite{KallinRPP2016}. The appearance of such non-singular vortices is an appealing hypothesis to explain the weakening of the pinning force and the existence of a vortex liquid state.

Both effects, the appearance of a "vortex liquid state" at high fields in UCoGe (pointed out by the crossing of bulk and resistive \Hc and by the narrowing of the resistive transition), as well as a possible change of the superconducting order parameter symmetry (detected by the abrupt increase of the residual thermal conductivity), whether they are related or not, show that the physics of $p$-wave ferromagnetic superconductors in transverse fields is extremely rich. They are worth further theoretical and experimental investigations, of the superconducting phase diagram, which could reveal new field-induced phase transitions, and of the vortices, which could display non-singular vortex cores, long predicted but only observed in superfluid $^3$He \cite{LounasmaaPNAS1999,MineevNature1986} up to now. 

\section {Acknowledgements:}
We are particularly thankful to M. E. Zhitomirsky for enlightening discussions on non singular vortices, as well as to V. Mineev, G. Knebel, A. Pourret, and D. Braithwaite.
	The work was supported by the ERC NewHeavyFermions, the ANR grant SINUS, and KAKENHI (JP15H05884, JP15H05882, JP15K21732, JP16H04006, JP15H05745).


\clearpage
\section{Supplemental Material}

\vspace{5\baselineskip}

\begin{figure}[!h]
\begin{center}
\includegraphics[width = 0.9\columnwidth]{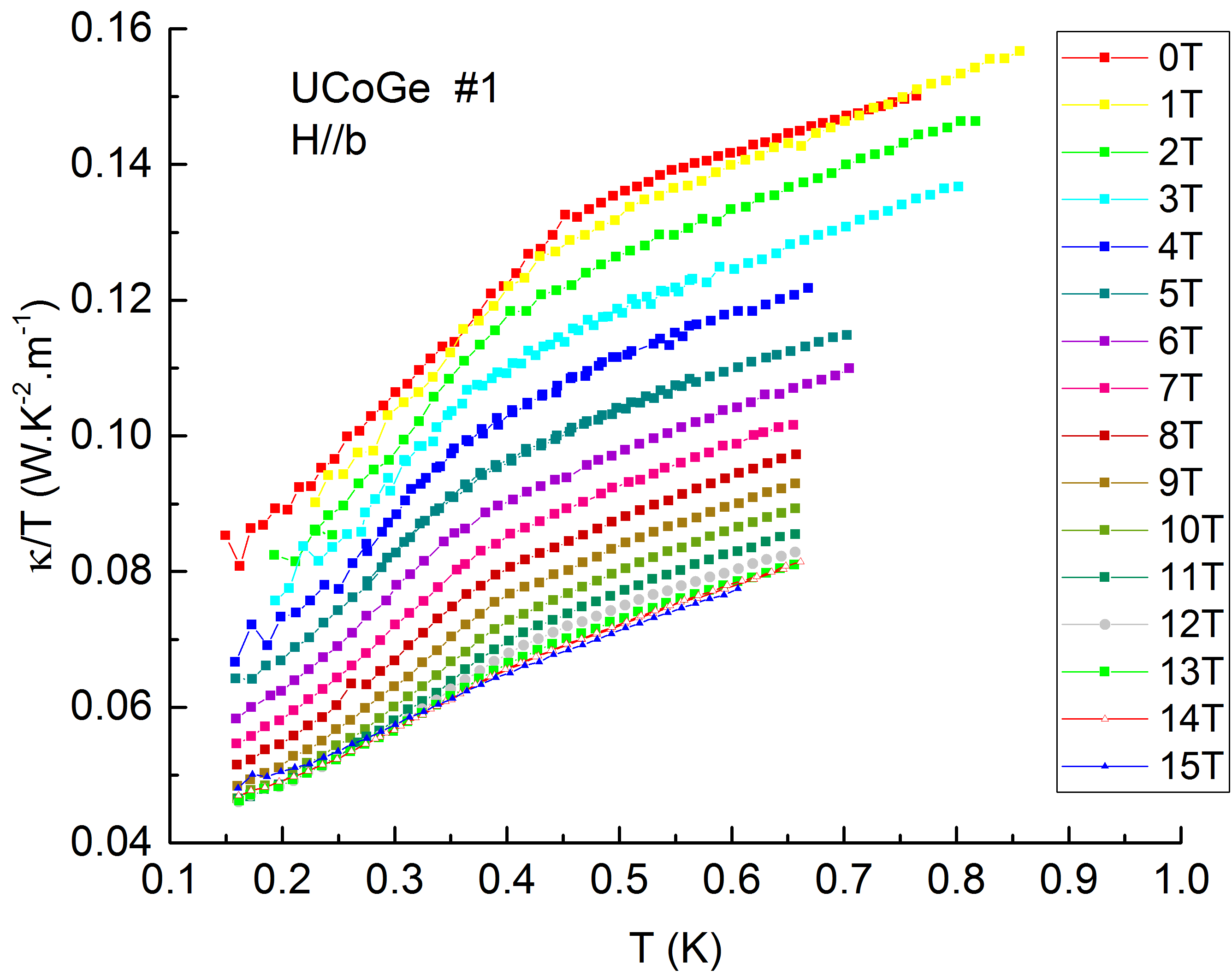}
\caption{\label{fig-kappa_sc16} Raw thermal conductivity data as a function of temperature measured on UCoGe sample $\#1$ (RRR=16), at different fields from 0T (top) to 15T (bottom) for \bH//\xb, with the heat current along the sample \xc axis.
}
\end{center}
\end{figure}

\begin{figure}[!t]
\begin{center}
\includegraphics[width = 0.9\columnwidth]{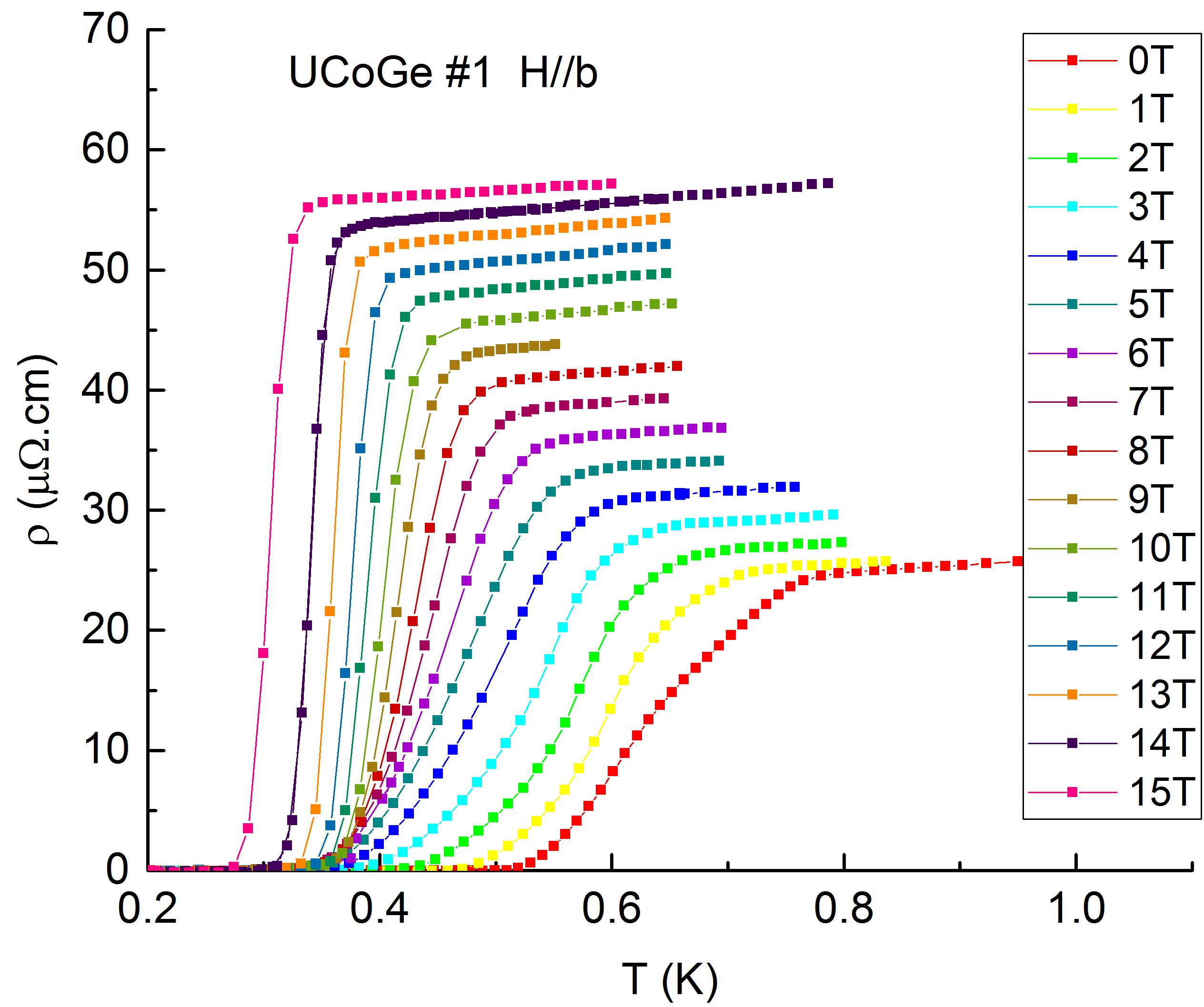}
\caption{\label{fig-rho_sc16} Resistivity as a function of temperature measured on UCoGe sample $\#1$ (RRR=16), at different fields from 0T (right) to 15T (left) for \bH//\xb, with the electrical current along the sample \xc axis.
}
\end{center}
\end{figure}


\begin{figure}[!b]
\begin{center}
\includegraphics[width = 0.9\columnwidth]{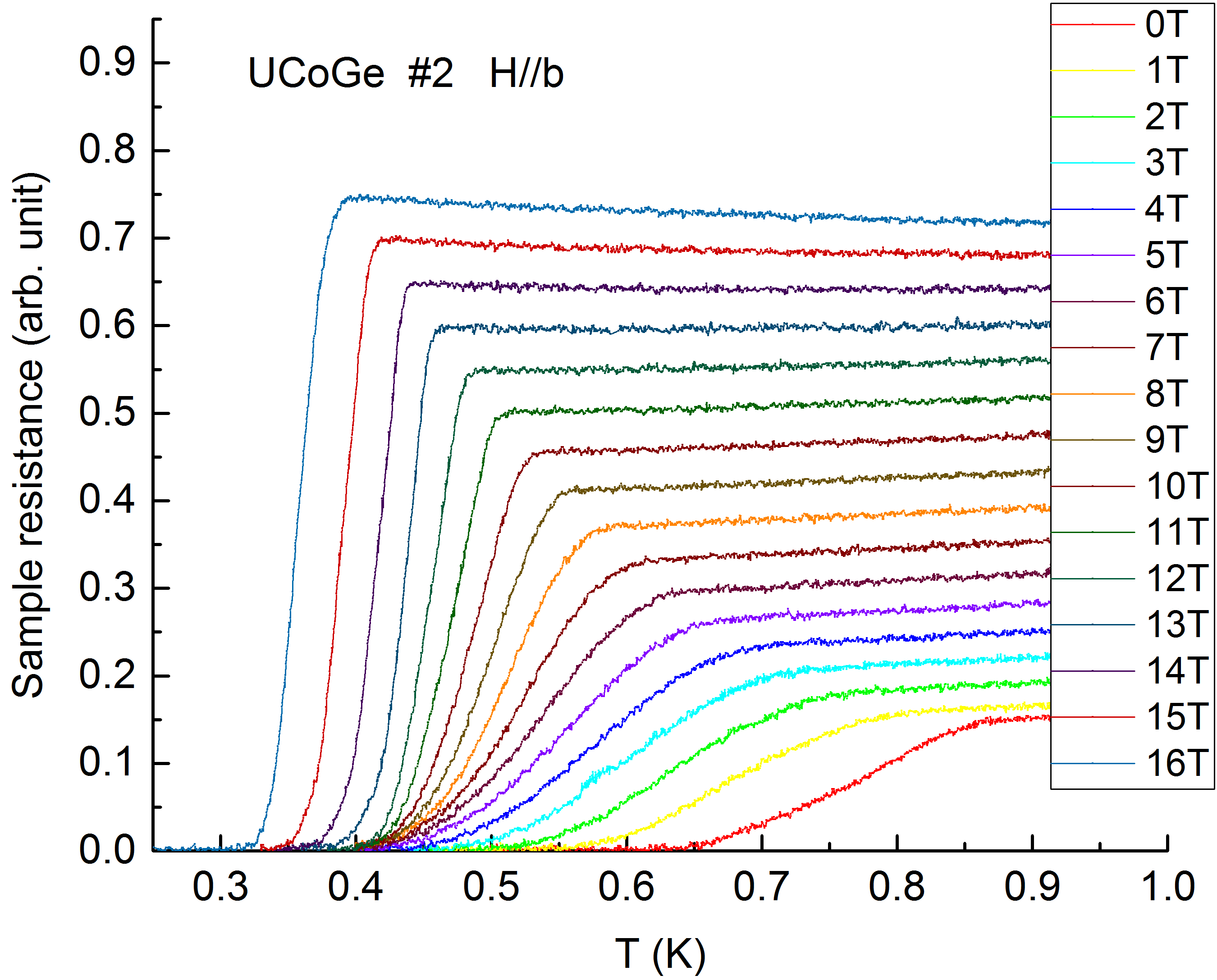}
\caption{\label{fig-rho_dai2} Resistivity as a function of temperature measured on UCoGe sample $\#2$ (RRR=35), at different fields from 0T (right) to 16T (left) for \bH//\xb, with the electrical current along the sample \xc axis.
}
\end{center}
\end{figure}

\end{document}